\documentclass[amsmath,amssymb,article,aps,prl,twocolumn,longbibliography]{revtex4-1}

\usepackage[latin1]{inputenc}
\usepackage{graphicx}
\usepackage{dcolumn}
\usepackage{bm}
\usepackage{color}
\newcommand{\ket}[1]{\left|#1\right\rangle}
\newcommand{\bra}[1]{\left\langle#1\right|}

\DeclareMathOperator{\tr}{tr}
\begin{document}

\title{Simple quantum password checking}
\author{Juan Carlos Garc\'ia-Escart\'in}
\email{juagar@tel.uva.es}
\author{Pedro Chamorro-Posada}
\affiliation{Dpto. de Teor\'ia de la Se\~{n}al y Comunicaciones. ETSI de Telecomunicaci\'on. Universidad de Valladolid. Campus Miguel Delibes. Paseo Bel\'en 15. 47011 Valladolid. Spain.}
\date{\today}
\begin{abstract}
We present a quantum password checking protocol where secrecy is protected by the laws of quantum mechanics. The passwords are encoded in quantum systems that can be compared but have a dimension too small to allow reading the encoded bits. We study the protocol under different replay attacks and show it is robust even for poorly chosen passwords. 
\end{abstract}

\pacs{42.50.Ex, 42.50.Tx, 03.67.Lx}
\maketitle
The password checking problem appears when two sides who share some common information want to verify the other side also knows it. We consider two parties, Alice and Bob, who share an $m$-bit string $p$ (the password). There can also be an eavesdropper, Eve, trying to either learn the password or to impersonate Alice or Bob. For simplicity, we describe the case where Alice tries to prove her identity to Bob. Everything is symmetric for the converse case. 

Imagine Alice and Bob can only communicate through an insecure classical channel. Alice could send her password in the clear, but Eve could read it undetected and later pose as Alice. A widespread solution is the use of hash functions. A hash, or one-way, function $H(x)$ takes a string $x$ into an output string of a fixed size that appears to be random. The function $H(x)$ should be easy to compute, but difficult to invert. The functions are chosen so that finding $x=H^{-1}(H(x))$ from $H(x)$ is computationally hard. If Alice sends $H(p)$, Eve could still copy the string and impersonate Alice, but, at least, the password is protected. If the password has some value, for instance, Alice uses it in other places, this is a small improvement. A greater advantage comes if Alice and Bob use hash chains. Alice can send a string $H^z(p)$ which results from applying the hash function $z$ times, first to $p$ and then to resulting hash of the previous steps. Bob can keep a record of the $z$s used in previous identification rounds and keep asking for smaller values of $z$. Eve can capture $H^z(p)$, but she is not able to produce $H^y(p)$ for $y<z$. This is a solution in classical networks \cite{Lam81}, but it is still vulnerable to dictionary attacks in which Eve compiles a list of the most common passwords and pre-computes their hashes and the hashes of their hashes up to a certain depth. If she captures a passing string $H^z(p)$, she can look it up in her table and find out the original password \cite{Kle90}.

There are password checking protocols, like SPEKE \cite{Jab96,Mac01}, which are based on Zero Knowledge Proofs \cite{GMR89} and avoid dictionary attacks from eavesdroppers (but not for a dishonest side \cite{Jab97}). Similary, the SSH protocol favours the use of public key cryptography coupled with challenge-response authentication. These systems are based on the computational difficulty of problems like factoring or the discrete logarithm, which, while robust against present technology, are not guaranteed to be hard problems and could, indeed, be broken with quantum computers \cite{Sho97}.

In this paper, we propose an alternative password checking protocol where security is derived from the laws of quantum mechanics. The protocol can resist replay attacks and is robust against dictionary attacks, even for poorly chosen passwords. 

Previous quantum solutions to this problem either use entangled states \cite{Bar99,BCG02,SLL01,LZ06,HO09} or are built on top of the quantum key distribution protocol BB84 \cite{BB84,DHH99,LBK00,DFL09}. Our protocol is instead based on quantum fingerprinting \cite{BCW01}. It encodes data in a quantum system too small to allow full recovery but which allows state comparison. However, our encoding focuses on security rather than in communication savings. In that respect, it is more similar to quantum public-key cryptography systems \cite{Nik08,Nik08b,NI09,IM09,IM11}.

We only need three primitives, quantum state preparation, random number generation and quantum state comparison. Quantum state comparison is performed with the SWAP test used in quantum fingerprinting. The test can detect different states. If we have two input states with density matrices $\rho$ and $\sigma$, the test fails and the states are proved to be different with probability $\frac{1-\tr (\rho \sigma)}{2}$. If the states pass the test, with probability $\frac{1+\tr (\rho \sigma)}{2}$, they can still be different. If we have many copies of the states, we can determine whether they are equal or different with high probability if they are different enough ($\tr(\rho\sigma)$ is not very close to 1). 

In our protocol, we apply the SWAP test on passwords that are encoded in symmetric states \cite{BKM97} of dimension $D=2^d$. We define symmetric states $\ket{\Phi_j}$ with $j=0,\ldots,N-1$ so that
\begin{equation}
\ket{\Phi_j}=\frac{1}{\sqrt{D}}\sum_{l=0}^{D-1}e^{i\frac{2\pi j l}{N}}\ket{l}.
\label{symm}
\end{equation}
We take states which encode both the password and a random bit string $r_i$ in a symmetric state $\ket{\psi_p^{r_i}}=\ket{\Phi_{H(p || r_i)}}$ with an index $j$ equal to the integer whose binary representation is the output of the hash function $H$ with an input that is the concatenation of the binary strings representing the password $p$ and the random string $r_i$. We work with a hash function $H(x)$ with $n$ bits of output so that there are $N=2^n$ possible password states that represent the $M=2^m$ possible passwords in a compressed state space.

A useful property of this encoding is that the mixed state that comes from choosing at random with the same probability any symmetric state $\rho_j=\ket{\Phi_j}\bra{\Phi_j}$ is the maximally mixed state as
\begin{multline}
\frac{1}{N}\sum_{j=0}^{N-1}\rho_j=\frac{1}{N}\frac{1}{D}\sum_{j=0}^{N-1}\sum_{l=0}^{D-1}\sum_{m=0}^{D-1}e^{i\frac{2\pi j (l-m)}{N}}\ket{l}\bra{m}=\\
\frac{1}{D}\sum_{l,m}\frac{1}{N}\sum_j (e^{i\frac{2\pi  (l-m)}{N}})^j\ket{l}\bra{m}=\frac{1}{D}\sum_{l}\ket{l}\bra{l}=\frac{I}{D},
\label{compMIX}
\end{multline}
using that the geometric sum of the $ e^{i\frac{2\pi j (l-m)}{N}}$ terms is only different from $0$ if $l=m$, when it is $N$.

With all these elements we can give the password checking protocol. We describe the case where Alice proves her identity to Bob and both share a password $p$ of $m$ bits that is hashed to produce $n$ bits. 
\newline

$\bullet$ Repeat up to $s$ times:

\begin{enumerate}
\item Alice and Bob generate jointly a random string $r_i$ of $m$ bits.
\item Alice prepares a state $\ket{\psi_p^{r_i}}$ in a quantum system of a dimension $D=2^d$ with $d\ll n$ and gives it to Bob.
\item Bob performs a quantum state comparison between the received state and a locally generated $\ket{\psi_p^{r_i}}$.
\begin{itemize}
\item[-] If the states are found to be different, Bob aborts the protocol.
\item[-] If the states pass the test, we repeat steps 1 to 3 with new random strings $r_i$ until we have $s$ positive comparisons.        
\end{itemize} 
\end{enumerate}

For simplicity we choose random strings of $m$ bits, but any bit number sufficiently larger than $n$ can be used. Alice and Bob can each generate random strings if they input orthogonal states into the SWAP test. They have probability $1/2$ of passing the test, which could be identified with a $1$ bit, and probability $1/2$ of failing (the bit is $0$). The joint random string can be produced by the XOR of Alice's and Bob's strings if they can guarantee the bits are simultaneously produced. For our purposes, they can also produce $r_i$ taking one bit from Alice and one from Bob. As long as one of the sides is honest, the hash function will introduce enough randomization even if only half the bits of $r_i$ are random. For a higher security, Alice and Bob can use the classical protocol of Damg{\aa}rd and Luneman, which offers unconditional security for one side and computational security, even against quantum computers, for the other side \cite{DL09}. Alternatively, they can use the unconditionally secure relativistic coin flipping protocol of Kent \cite{Ken99b} or strings whose randomness is certified by Bell's theorem \cite{PAM10}.

The basic concept of the protocol can be explained with a simple analogy. We can picture the quantum state as a piece of paper with a limited size. If we try to write too many words, at some point we need to sacrifice legibility. The greater the number of words, the more difficult it becomes to make sense of what is written. However, it is still possible to compare two sheets of paper. A quick glance suffices to tell, with good accuracy, whether the contents are equal or not. If the contents are almost equal, there will be a good chance of giving a false positive, but completely different texts, say a page full of the letter ``a'' and one only with numbers, can be told apart with high confidence. 

In this protocol, the password is protected by the limitations of the quantum encoding. We have encoded too many bits, $n$, into a too small state space. From the Holevo bound we know we cannot reliably recover more than $d$ bits from a $D=2^d$-dimensional quantum state \cite{Hol73}. From Nayak's bound we also know that for $d<<n$, we also have a small probability of recovering only some of those bits \cite{Nay99}. We use a stronger result from Ben-Aroya, Regev and de Wolf \cite{BRW08} which limits $k$-out-of-$n$ encodings that allow to recover $k$ bits from the total $n$ bits of a string. If $\frac{d}{n}<\frac{1}{2\ln 2}\approx 0.72$, the probability of recovering $k$ bits is exponentially small in $k$. For a small enough $\frac{d}{n}$, the probability tends to $2^{-k}$, which is as likely as guessing the bits by chance. 

If we choose $sd\ll n$, Eve would not learn anything from the password, even if she captured all the states in a full protocol exchange. Usually, Alice and Bob must both identify to the other side. In those cases, instead of first performing the $s$ stages of Alice authenticating to Bob and the $s$ stages of Bob convincing Alice, they should take turns for each stage. If Eve is on any of the sides, she would be detected early, before she can capture many states. The number of generated states before changing the password must be established so that the maximum captured states $c$ still satisfies $cd\ll n$. 

The $k$-out-of-$n$ bound also limits the amount of global information an attacker can learn from the password state. The useful information about any $k$ bits of $H(p||r_i)$ must be, at most, exponentially small in $k$. Otherwise, Alice and Bob could devise a code that groups the states with the different values of a global property, such as the parity of all the bits or the AND function of a group of them, and use the attack as a way to send one bit of information with a probability better than the limit close to 0.5 given by the bound.

If we have $M$ possible combinations of parameters, the average probability of passing the SWAP test is related to the average fidelity $\bar{F}$ of all the fidelities $F_k=\tr (\rho_t \rho_k)$ between the trial state $\rho_t$ of an attacker who does not know $p$ and the legitimate pure state $\rho_k$ which depends on the considered parameters. We have,
\begin{equation}
\bar{P}=\frac{1}{M}\sum_{k=0}^{M-1}\frac{1+F_k}{2}=\frac{1+\frac{1}{M}\sum_k F_k}{2}=\frac{1+\bar{F}}{2}.
\end{equation}
In the rest of the paper we show an attacker who doesn't know anything about $p$ can only produce states with an average fidelity $\bar{F}=\frac{1}{D}$. If $D$ is high enough, the probability of passing the $s$ stages of the protocol and fooling Bob into accepting a false Alice is 
\begin{equation}
\left(\frac{1}{2}+\frac{1}{2D}\right)^s\approx 2^{-s}+2^{-s}\frac{s}{D},
\end{equation}
which can be made arbitrarily small. 

There are two main attacks we must avoid. The first is a {\bf random state attack}. We can consider Eve intends to fool Bob with a trial state $\rho_t$ that is constant for all attempts. She will try to maximize her probability of passing the test for all the possible passwords. The average fidelity for this trial state is
\begin{equation}
\bar{F}=\frac{1}{M}\sum_{p=0}^{M-1}\tr(\rho_t\rho_p^{r_i})=\tr\left(\rho_t\frac{1}{M}\sum_{p=0}^{M-1}\rho_p^{r_i}\right)
\end{equation}
for the constant, known $r_i$ of each stage. When we consider all the possible fixed $r_i$, the value of an ideal hash function $H(p||r_i)$ is evenly distributed through all the integers from $0$ to $N-1$ with the different values of $p$. When the number of the hash output bits $n$ is sufficiently smaller than the number of the password bits $m$, the sum gives a uniform average on all the possible symmetric states and the result is close to the completely mixed state of Equation (\ref{compMIX}). The average fidelity can then be approximated by
\begin{equation}
\bar{F}=\tr\left(\rho_t\frac{I}{D}\right)=\frac{1}{D}\tr(\rho_t)=\frac{1}{D}
\end{equation}
if $\rho_t$ is pure, or smaller for mixed trial states. Essentially, $p$ randomizes the state. The attacker sees a maximally mixed state and cannot do better than random guessing. 

There is a second important family of attacks. Eve can try a {\bf replay attack} in which she captures and stores previous legitimate interchanges between Alice and Bob and uses the captured states to impersonate Alice. The random strings $r_i$ prevent Eve from directly using her stored states, but she can try to modify them to fool Bob.

The simplest case is one in which Eve captures a state $\rho_p^{r_1}$ and wants to produce a state as close to the next state in the protocol $\rho_p^{r_2}$ as possible. Eve can perform any allowed operation on the captured state. The most general transformation allowed by quantum mechanics is a completely positive map $\mathcal{E}(\rho_p^{r_1})=\tr_E(U(\rho_p^{r_1}\otimes \rho_{anc})U^\dag)$ which combines the effects of adding an ancillary system $\rho_{anc}$, performing a unitary operation $U$ in the larger state space of $\rho_p^{r_1}\otimes \rho_{anc}$ and then presenting Bob with a subsystem of the resulting state, represented by the partial trace $\tr_E$ \cite{Kra83}. Eve wants to maximize  her fidelity 
\begin{equation}
\bar{F}=\frac{1}{M^3}\sum_{p=0}^{M-1}\sum_{r_1=0}^{M-1}\sum_{r_2=0}^{M-1}\tr(\mathcal{E}(\rho_p^{r_1})\rho_p^{r_2})
\label{avF}
\end{equation}
for any combination of $p$, $r_1$ and $r_2$. 

We use a result by Fiur\'a\v{s}ek to bound this fidelity \cite{Fiu01}. For input states $\ket{\psi_{in}(\vec{x})}$ chosen from a set of parameters $\vec{x}$, the best approximation to outputs $\ket{\psi_{out}(\vec{x})}$ is given by the average fidelity
\begin{equation}
\bar{F}=\int  \bra{\psi_{out}(\vec{x})} \mathcal{E}(\ket{\psi_{in}(\vec{x})}\bra{\psi_{in}(\vec{x})})\ket{\psi_{out}(\vec{x})}\,d\vec{x},
\end{equation}
which can be optimized using Lagrange multipliers. For our discrete version of the average fidelity shown in Equation (\ref{avF}), we need to look at operator
\begin{equation}
\hat{R}=\frac{1}{M^3}\sum_{p=0}^{M-1}\sum_{r_1=0}^{M-1}\sum_{r_2=0}^{M-1}(\rho_p^{r_1}\otimes \rho_p^{r_2}).
\end{equation}
For any fixed $p$, the sum in $r_1$ and $r_2$ has outputs $H(p||r_1)$ and $H(p|| r_2)$ in all the values of $0$ to $N-1$ and Equation (\ref{compMIX}) is a good approximation to the final state so that
\begin{equation}
\hat{R}=\frac{1}{M^2}\sum_{p}\sum_{r_1}\left(\rho_p^{r_1}\otimes \frac{I}{D}\right)=\frac{1}{M}\sum_{p}\left(\frac{I}{D}\otimes \frac{I}{D}\right)=\frac{I}{D^2}.
\end{equation}
Fidelity $\bar{F}$ is bounded by the dimension of the input Hilbert space $\dim \mathcal{H}$ and the largest eigenvalue $R_{max}$ of $\hat{R}$ \cite{Fiu01}. In our case, $\frac{I}{D^2}$ has eigenvalues $\frac{1}{D^2}$ and 
\begin{equation}
\bar{F}\le \dim \mathcal{H} R_{max}=D\frac{1}{D^2}=\frac{1}{D}.
\end{equation}
We can extend the results to the case where Eve captures up to $c$ states $\rho_p^{r_1}, \rho_p^{r_2},\ldots, \rho_p^{r_c}$ and wants to approximate the next state in the sequence $\rho_p^{c+1}$. Now
\begin{equation}
\bar{F}=\frac{1}{M^{c+2}}\sum_{p,r_1,\ldots,r_c,r_{c+1}}\tr(\mathcal{E}(\rho_p^{r_1}\otimes \rho_p^{r_2} \otimes \cdots \otimes \rho_p^{r_{c}})\rho_p^{r_{c+1}})
\label{avCapt}
\end{equation}
and
\begin{equation}
\hat{R}=\frac{1}{M^{c+2}}\sum_{p,r_1,\ldots,r_c,r_{c+1}} (\rho_p^{r_1}\otimes \rho_p^{r_2} \otimes \cdots \otimes \rho_p^{r_{c}}\otimes \rho_p^{r_{c+1}}).
\end{equation}
We can proceed in a way similar to the single captured state scenario and show
\begin{equation}
\hat{R}=\frac{I}{D^{c+1}}
\end{equation}
and
\begin{equation}
\bar{F}\le \dim \mathcal{H}^{\otimes c} R_{max}=D^c \frac{1}{D^{c+1}}=\frac{1}{D}.
\end{equation}
There are two caveats. First, the bound breaks down after a number of captured states. We assume Eve doesn't know a single bit of $p$. If she did, the average of Equation (\ref{avCapt}) is not the quantity to maximize and the proof is no longer valid. The bound holds as long as $cd\ll n$ so that Eve cannot guess the bits from $p$. We already keep this bound in order to protect the password. 

There might be insecure particular cases, but, for random $r_i$s, they must be rare on average. After $s$ stages, the probability of fooling the system still tends to $2^{-s}+2^{-s}\frac{s}{D}$. Bob can also keep a list of all the previously used $r_i$s to avoid repetition or weak special cases. 

These bounds show the protocol is resistant to eavesdroppers that try to learn the password or perform a replay attack. The protocol offers and additional layer of security beyond that of systems based on computational complexity. Security is based on physical limitations. The protocol protects against adversaries with a quantum computer and is independent of technological assumptions. A hash $H(x)$ which is difficult to invert with present technology could be broken with better computers. This has happened with algorithms like MD5 \cite{WY05}. Similarly, the systems that use public key cryptography are usually optimized for speed and could be broken with better computers. In our protocol, if we choose a small enough dimension $D$, the password is unreadable even for future technologies. 

Using quantum systems also gives additional protection from dictionary attacks. If we choose a bad password $b$ from a reduced set of possible passwords of size $B$ so that $D\ll B< 2^m$, Eve still cannot look up the captured states in a pre-computed table. If she could find the password, she would have an encoding that allowed to squeeze more than $d$ bits into a $D$-dimensional state. 

A bad password choice would not reduce the security of the system as long as $B$ is sufficiently larger than $N$. If we suspect passwords are chosen from a smaller set, like a small dictionary of English words, our proofs are no longer valid and there could exist advanced impersonation attacks. If passwords are chosen freely, it could be safer to send a hash $q=H'(b)$ that takes the passwords into a string with a smaller size $n'$ which is small enough to make bad passwords produce randomly distributed $q$ strings and, at the same time, is large enough to make $cd\ll n'$. This reduces the number of password reuses, but preserves the properties of the original scheme.

The protocol has been given in a general form with three simple primitives that do not require a full quantum computer. A practical realization of the protocol can be deployed on top of existing systems as an additional measure of security. There is a relatively simple implementation with optical primitives where state preparation is feasible and the SWAP test can be performed with the Hong-Ou-Mandel effect \cite{SvE11,GC13}. A detailed implementation will be presented elsewhere.

\begin{acknowledgments}
This work has been funded by MICINN project TEC2010-21303-C04-04.
\end{acknowledgments}
\newcommand{\noopsort}[1]{} \newcommand{\printfirst}[2]{#1}
  \newcommand{\singleletter}[1]{#1} \newcommand{\switchargs}[2]{#2#1}
\end{document}